\documentclass[12pt]{article}
\usepackage[left=2.5cm,top=2.50cm,right=2.5cm,bottom=2.50cm]{geometry}
\usepackage{mathrsfs}
\usepackage{amsmath,amssymb,latexsym,color,cancel,graphicx,bbm,colortbl}
\usepackage[english]{babel}
\usepackage[latin1]{inputenc}
\usepackage{ragged2e}
\usepackage{cite}
\DeclareMathAlphabet{\mathpzc}{OT1}{pzc}{m}{it}
\begin{document}
\date{}

\title{Algebraic Approach and Coherent States for the Modified Dirac Oscillator in Curved
Spacetime with Spin and Pseudospin Symmetries}
\author{M. Salazar-Ram\'irez$^{a}$\footnote{{\it E-mail address:} escomphysics@gmail.com}, D. Ojeda-Guill\'en$^{a}$,  J.A. Martínez-Nuño$^{a}$, \\ R.I. Ramírez-Espinoza$^{a}$} \maketitle

\begin{minipage}{0.9\textwidth}
\small $^{a}$ Escuela Superior de C\'omputo, Instituto Polit\'ecnico Nacional,
Av. Juan de Dios B\'atiz esq. Av. Miguel Oth\'on de Mendiz\'abal, Col. Lindavista,
Alc. Gustavo A. Madero, C.P. 07738, Ciudad de M\'exico, Mexico.\\
\end{minipage}

\begin{abstract}
In this article we investigate and solve exactly the modified Dirac oscillator in curved spacetime with spin and pseudospin symmetries through an algebraic approach. By focusing on the radial part of this problem, we use the Schr\"odinger factorization method to show that this problem possesses an $SU(1,1)$ symmetry. This symmetry allowed us to obtain the wave functions and their corresponding energy spectrum. From these results, we calculate the radial coherent states of the modified Dirac oscillator and their temporal evolution in the spin and pseudospin limits, respectively.
\end{abstract}


\section{Introduction}
Since the german physicists Otto Stern and Walther Gerlach published their work on spin in 1922, numerous studies related to this discovery have been conducted. With this experiment, the authors conclude that electrons, in addition to having orbital angular momentum, also possess a quantized intrinsic angular momentum known as spin\cite{Ger}.

Approximately five decades ago, the concept of pseudospin symmetry was introduced as a result of observations concerning the quasidegeneracy of doublets of single nucleons exhibiting non-relativistic quantum numbers $\left(n,l,j=l+1/2\right)$ and $\left(n-1,l+2,j=l+3/2\right)$, where $n$, $l$ and $j$ are single-nucleon radial, orbital and total angular momentum quantum numbers respectively. The spin symmetry is present in the spectrum of mesons with a heavy quark and helps explain the absence of quarks spin-orbit splitting (spin doublets), observed in both heavy and light quarks \cite{Ari,Hec,ABE,AKB}.

In the case of pseudospin symmetry, it accounts for the degeneracies of certain shell-model orbitals in nuclei (pseudospin doublets). Moreover, it explains features not only related to deformation and superdeformation in nuclei but also establishes an effective scheme for the nuclear shell-model \cite{Gino,Dud,Trol}.

Ginocchio revealed in his works that the emergence of pseudospin symmetry in nuclei is attributed to a relativistic symmetry in the Dirac Hamiltonian \cite{Gino,Gino2,Gino3,Gino4}. In this series of works, the author demonstrates that spin symmetry occurs when the difference between the vector potential and the scalar potential is a constant, and pseudospin symmetry emerges when the sum of these two potentials is a constant. From the theory, it is well known that the pseudospin symmetry is exact under the condition $\frac{d\left(V(r)+S(r)\right)}{dr}=constant$, and spin symmetry occurs when $\frac{d\left(V(r)-S(r)\right)}{dr}=constant$.

Since the first publications on spin and pseudo-spin symmetry, the Dirac oscillator \cite{Ito,Cook,Moshi} under these conditions has been the subject of numerous studies, using a wide variety of potentials and methods to obtain the energy eigenvalue and the corresponding radial wave functions \cite{Chen,Lisboa,Gino5,Xu}. It is important to note that all these works were developed considering a flat space-time.

Analogously, the Dirac oscillator has also been studied in a curved spacetime, as can be seen in Refs. \cite{Lima1,Chen2,Oliv1}. In particular, in Ref. \cite{Lima1}, the authors studied a two-dimensional Dirac oscillator in cosmic string spacetime. They derive and solve the Dirac-Pauli equation in the limit of pseudospin and spin symmetries. Also, they analyze the presence of cylindrical symmetric scalar potentials and obtain analytical solutions for the resulting field equation \cite{Lima1}. In Ref.\cite{Chen2}, it is solved the radial problem of the Dirac oscillator in the cosmic string spacetime, using the Killingbeck potential plus isotonic oscillator potential in the limit of spin and pseudospin symmetries, through the functional Bethe ansatz method.

Recently, in Ref. \cite{Oliv2} the modified Dirac oscillator with spin and pseudo-spin symmetries in deformed nuclei has been investigated in depth. Both cases have been solved exactly and the energy spectrum has been obtained. In addition, the probability density and some energy spectra for both cases have been meticulously analyzed. Hence, the aim of this work is to study and solve exactly the modified Dirac oscillator in curved spacetime with spin and pseudospin symmetries from an algebraic point of view. We also calculate the radial coherent states and their time evolution.

This work is organized as it follows. In Section $2$, we provide a brief mathematical review of the modified Dirac oscillator in curved spacetime. Section $3$ is dedicated to derive the uncoupled second-order differential equations satisfied by the radial components by setting the spin limit. Using a purely algebraic approach, we obtain the energy spectrum and eigenfunctions by means of the theory of irreducible representations of the $su(1,1)$ Lie algebra. Then, we obtain the Perelomov $SU(1,1)$ radial coherent states for these eigenstates. In Section $4$, we obtain results analogous to those presented in Section $3$ for the pseudospin limit. Hence, we apply the condition for the pseudospin limit to appear and calculate its energy spectrum, eigenfunctions and radial coherent states. In Section $5$, we analyze the temporal evolution of these coherent states in both limits. Additionally, using our method, we obtain the energy spectrum without spin symmetry in flat spacetime. Finally, we provide some comments and conclusions regarding our work.

\section{A Review of the Dirac Equation in Curved Spacetime}

In this work we shall consider the following line element with spherical symmetry
\begin{equation}
ds^2=e^{2f(r)}dt^2-e^{2g(r)}dr^2-r^2d\theta^2-r^2\sin^2{\theta}d\phi^2,
\end{equation}
where $f(r)$ and $g(r)$ are arbitrary functions of the radial coordinate $r$. By setting $e^f=e^g=1+U(r)/c^2$, the line element in the curved spacetime is now given by \cite{Oliv2}
\begin{align}\label{eline}
ds^2=(1+U(r)/c^2)^2(dt^2-dr^2)-r^2d\theta^2-r^2sin^2\theta d\phi^2,
\end{align}
where $c$ is light speed and $U(r)$ is the scalar potential which contains all the information about the spacetime curvature. With the electromagnetic field given by $A_{\mu}=\left(V(r),cA(r),0,0\right)$, the Dirac Hamiltonian in atomic units is
\begin{align}\label{ham1}\nonumber
&\frac{H}{c}=-i\alpha^1_d\left(\partial_r+\frac{1}{r}+\frac{f'_{r}}{2}+\beta{A(r)}\right)-i\frac{e^f}{r}\left\lbrace\alpha^2_d\left(\partial_{\theta}+\frac{\cot{\theta}}{2}\right)+\frac{\alpha^3_d}{\sin\theta}\partial\phi\right\rbrace\\
&+c\beta e^f+\frac{V(r)}{c}.
\end{align}
In this expression $\alpha^i_d$ and $\beta$ are the Dirac matrices, and it was introduced the minimal coupling $p_0\rightarrow p_0-A_0/c$ and $\vec{p}\rightarrow\vec{p}-i\beta\vec{A}/c$, with $\vec{A}=(cA(r),0,0)$. Hence, by making the unitary transformation $H'=D(H/c)D^{\dag}$ and $\Psi'=D\Psi$ in terms of $D=\left(\bf{1_4}+\gamma^2\gamma^1+\gamma^1\gamma^3+\gamma^3\gamma^2\right)/2$ in equation (\ref{ham1}) we obtain \cite{Oliv2}
\begin{align}\nonumber
\frac{E}{c}\Psi'=&\left[-i\alpha^3\left(\partial_r+\frac{1}{r}+\frac{f'_r}{2}+\beta{A(r)}\right) -i\frac{e^f}{r} \left\lbrace\alpha^1\left(\partial_{\theta}+\frac{\cot\theta}{2}\right)+\frac{\alpha^2}{\sin\theta}\partial\phi\right\rbrace+ce^f\beta+\right.\\
&\left.\frac{V(r)}{c}\right]\Psi'.
\end{align}
Therefore, the $4$-component spinorial wave function can be written as
\begin{equation}
\Psi(r,\theta,\phi)=N\frac{e^{-f/2}}{r}\begin{Bmatrix}
R_{1\lambda}(r)\mathcal{Y}^{|m|j}_l\left(\theta,\phi\right)\\
R_{2\lambda}(r)\mathcal{Y}^{|m|j}_{l+1}\left(\theta,\phi\right)
\end{Bmatrix},
\end{equation}
where $\mathcal{Y}^{|m|j}$ are the spinorial functions given by
\begin{equation}
\mathcal{Y}^{j=l\pm{1/2},m}_l\left(\theta,\phi\right)= \frac{1}{\sqrt{2l+1}}\begin{Bmatrix}
\pm\sqrt{l\pm m+1/2}Y^{m-1/2}_l\left(\theta,\phi\right)\\
\sqrt{l\mp m+1/2}Y^{m+1/2}_l\left(\theta,\phi\right)
\end{Bmatrix}.
\end{equation}
Here, $l=j\mp\frac{1}{2}$, $Y^{m}_l$ are the spherical harmonics, and $R_{1\lambda}(r)$ and $R_{2\lambda}(r)$ satisfy the following radial equation \cite{Oliv2}
\begin{equation}\label{hamprinc}
\begin{pmatrix}
e^f+\alpha^2V(r)-\epsilon & -\alpha\left[\frac{d}{dr}-A(r)-\frac{\lambda}{r}e^f\right]\\\\
\alpha\left[\frac{d}{dr}+A(r)+\frac{\lambda}{r}e^f\right] & -e^f+\alpha^2V(r)-\epsilon
\end{pmatrix}
\begin{pmatrix}
R_{1\lambda}(r)\\\\
R_{2\lambda}(r)
\end{pmatrix}=0,
\end{equation}
where $\epsilon=E/c^2$ and $c=1/\alpha$, being $\alpha$ the fine structure constant. Now, we shall focus on the radial differential equations obtained from the expression (\ref{hamprinc}) and using $e^f=1+\alpha^2U(r)$ for particle and anti-particle ($R_{1\lambda}$ and $R_{2\lambda}$ respectively), we obtain
\begin{equation}\label{HAMDIC}
H_{\lambda}\Phi_{\lambda}=
\begin{pmatrix}
1+\alpha^2\Sigma(r) & -\alpha\left[\frac{d}{dr}-\frac{\lambda}{r}-\lambda\alpha^2\frac{U(r)}{r}-A(r)\right]\\
\alpha\left[\frac{d}{dr}+\frac{\lambda}{r}+\lambda\alpha^2\frac{U(r)}{r}+A(r)\right] & -1+\alpha^2\Delta(r)
\end{pmatrix}\Phi_{\lambda}=\epsilon\Phi_{\lambda}.
\end{equation}
In this expression, $\Phi_{\lambda}$ is the radial function given by
\begin{equation}
\Phi_{\lambda}=\begin{pmatrix}
R_{1\lambda}(r)\\
R_{2\lambda}(r)
\end{pmatrix},
\end{equation}
and $\Sigma(r)$, $\Delta(r)$ are explicitly defined in terms of the vector and scalar potentials $V(r)$ and $U(r)$ as
\begin{equation}
\Sigma(r)=V(r)+U(r), \hspace{0.5cm} \Delta(r)=V(r)-U(r).
\end{equation}
From equation (\ref{HAMDIC}) the coupled differential equations for $R_{1\lambda}(r)$ and $R_{2\lambda}(r)$ can be expressed by the following pair of equations \cite{Oliv2}
\begin{align}\label{eqdif1}
\left(1+\alpha^2\Sigma(r)-\epsilon\right)R_{1\lambda}(r)-\alpha\left(\frac{d}{dr}-\frac{\lambda}{r}-\lambda\alpha^2\frac{U(r)}{r}-A(r)\right)R_{2\lambda}(r)=0,\\\label{eqdif2}
\alpha\left(\frac{d}{dr}+\frac{\lambda}{r}+\lambda\alpha^2\frac{U(r)}{r}-A(r)\right)R_{1\lambda}(r)+\left(-1+\alpha^2\Delta(r)-\epsilon\right)R_{2\lambda}(r)=0.
\end{align}

\section{The spin limit $\Delta(r)=0, V(r)=U(r), \Sigma(r)=2U(r)$}

The spin limit occurs when the scalar and vector potentials are equal, that is, when $\Delta(r)=0$. Under this limit, from equations (\ref{eqdif1}) and (\ref{eqdif2}) we can write the decoupled differential equation for $R_{1\lambda}(r)$ as \cite{Oliv2}
\begin{align}\label{eqdif3}\nonumber
&\left[\frac{d^2}{dr^2} +\frac{d}{dr}\left(\frac{\lambda}{y}+\lambda\alpha^2\frac{U(r)}{r}+A(r)\right)-\left(\frac{\lambda}{r}+\lambda\alpha^2\frac{U(r)}{r}+A(r)\right)^2-\right.\\
&\left.-\frac{\left(1-\epsilon+\alpha^2\Sigma(r)\right)\left(1+\epsilon\right)}{\alpha^2}\right]R_{1\lambda}(r)=0.
\end{align}
Now, we will consider a quadratic radial potential $U={\mu}r^2$ and the tensor potential as $A(r)=wr+A/r$. From these conditions, the effective tensor potential is $A/r+\beta_1 r$, where $\beta_1=\omega+\alpha^2\lambda\mu$ is the partial oscillation frequency. Thus, we obtain from the expression (\ref{eqdif3})
\begin{equation}\label{eqdif4}
\left[\frac{d^2}{dr^2}-\frac{\lambda\left(\lambda+1\right)+A\left(2\lambda+A+1\right)}{r^2}-\delta^2r^2+\beta+\frac{\epsilon^2-1}{\alpha^2}\right]R_{1\lambda}(r)=0,
\end{equation}
where the total oscillation frequency is given by $\delta=\sqrt{\beta_1^2+2\mu(1+\epsilon)}$ and $\beta=\beta_1(1-2\lambda-2A)$. By making the substitution $\left(\lambda+1\right)=l\left(l+1\right)$, $x={\delta}r^2$ and $R_{1\lambda}(r)=x^{-\frac{1}{4}}\Phi$ the equation (\ref{eqdif4}) becomes \cite{Oliv2}
\begin{equation}\label{equdiffac}
\left[-x\frac{d^2}{dx^2}+\left(\frac{l\left(l+1\right)+A\left(2\lambda+A+1\right)}{4}-\frac{3}{16}\right)\frac{1}{x}+\frac{x}{4}\right]\Phi=\left(\frac{\beta}{4\delta}+\frac{\epsilon^2-1}{4\delta\alpha^2}\right)\Phi.
\end{equation}

From equation (\ref{equdiffac}) we can construct the $su(1,1)$ algebra generators for the radial function $\Phi$ by applying the Schr\"odinger factorization. Thus, if we
propose
\begin{equation}\label{factosch}
\left(x\frac{d}{dx}+ax+b\right)\left(-x\frac{d}{dx}+cx+f\right)\Phi=g\Phi,
\end{equation}
we can expand this expression and compare it with the equation (\ref{equdiffac}) to obtain that the constants $a,c,f,b,g$ are
\begin{align}
a=&\pm\frac{1}{2},\hspace{0.5cm} c=\pm\frac{1}{2}, \hspace{0.5cm} f=\mp\left(\frac{\beta}{4\delta}+\frac{\epsilon^2-1}{4\delta\alpha^2}\right),\hspace{0.5cm}b=\mp\left(\frac{\beta}{4\delta}+\frac{\epsilon^2-1}{4\delta\alpha^2}\right)-1,\\
g=&\left[-\frac{l\left(l+1\right)+A\left(2\lambda+A+1\right)}{4}+\frac{3}{16}\right]+\left[\mp\left(\frac{\beta}{4\delta}-\frac{\epsilon^2-1}{4\delta\alpha^2}\right)-1\right]\left[\mp\left(\frac{\beta}{4\delta}-\frac{\epsilon^2-1}{4\delta\alpha^2}\right)\right].
\end{align}
Thus, the differential equation for $\Phi$ can be factorized as
\begin{equation}
\left[\mathscr{Z}_{\mp}\mp1\right]\mathscr{Z}_{\pm}=\Pi+\Gamma\left(\Gamma\pm1\right),
\end{equation}
where
\begin{equation}
\Pi=\left[-\frac{l\left(l+1\right)+A\left(2\lambda+A+1\right)}{4}+\frac{3}{16}\right], \hspace{0.5cm}\Gamma=\left(\frac{\beta}{4\delta}-\frac{\epsilon^2-1}{4\delta\alpha^2}\right),
\end{equation}
and the Schr\"odinger operators are given by
\begin{align}
\mathscr{Z}_{\pm}=\mp x\frac{d}{dx}+\frac{x}{2}-\left(\frac{\beta}{4\delta}-\frac{\epsilon^2-1}{4\delta\alpha^2}\right).
\end{align}
From these results, we can introduce a new pair of operators given by
\begin{equation}\label{OPT}
\mathscr{T}_{\pm}=\mp x\frac{d}{dx}+\frac{x}{2}-\mathscr{B}_3,
\end{equation}
with $\mathscr{B}_3$ being an operator defined from equation (\ref{equdiffac}) as
\begin{equation}\label{OPER3}
\mathscr{B}_3=\left[-x\frac{d^2}{dx^2}+\left(\frac{l\left(l+1\right)+A\left(2\lambda+A+1\right)}{4}-\frac{3}{16}\right)\frac{1}{x}+\frac{x}{4}\right]\Phi=\left(\frac{\beta}{4\delta}-\frac{\epsilon^2-1}{4\delta\alpha^2}\right)\Phi.
\end{equation}
It is straightforward to demonstrate that the operators given by equations (\ref{OPT}) and (\ref{OPER3}) close the $su(1,1)$ Lie algebra
\begin{equation}
\left[\mathscr{B}_3, \mathscr{T}_{\pm}\right]=\mp\mathscr{T}_{\pm},\hspace{0.5cm} \left[\mathscr{T}_-,\mathscr{T}_+\right]=2\mathscr{B}_3.
\end{equation}

Now, let us obtain the energy spectrum for the spin limit using the theory of unitary irreducible representations of the $su(1, 1)$ Lie algebra\cite{MSR1}
\begin{align}\label{tercerop}
\mathscr{B}_3|k,n\rangle&=(k+n)|k,n\rangle,\\\label{casi}
\mathscr{C}^2&=-\mathscr{T}_+\mathscr{T}_-+\mathscr{B}_3(\mathscr{B}_3-1)=k(k-1),
\end{align}
where $\mathscr{C}$ is the Casimir operator. Thus, from equations (\ref{OPT}), (\ref{OPER3}) and (\ref{casi}) the quantum number $k_S$ $\left(k\rightarrow k_S\right)$ in the spin limit can be written as
\begin{align}\label{numk}
k_{S}=\frac{1}{2}\pm\frac{\sqrt{1+4A^2+8A\lambda+4A+4l+4l^2}}{4}.
\end{align}
The equations (\ref{OPER3}) and (\ref{tercerop}) let us write the following expression
\begin{equation}\label{numkann}
k_{S}+n=\left[\frac{\beta}{4\delta}+\frac{\epsilon^2-1}{4\delta\alpha^2}\right].
\end{equation}
From this equation we obtain the fourth-degree polynomial which will allow us to calculate the energy spectrum by solving the polynomial
\begin{equation}\label{ecuener}
\epsilon^4+p\epsilon^2+q\epsilon+r=0,
\end{equation}
where
\begin{align}\label{var1}
p=&2a, \hspace{1.0cm} q=\left(-2{\mu}b^2\right),\hspace{1.0cm} r=a^2-b^2c,\hspace{1.0cm} a=-1+\beta\alpha^2, \\\label{var2}
b=&4\alpha^2\left(k_{s}+n\right),\hspace{1.0cm} c=\left(2\mu+\left(w+\alpha^2\lambda\mu\right)^2\right).
\end{align}
By using Ferrari's solution \cite{Ferr}, we obtain
\begin{align}\label{especspin1}
\epsilon=\frac{\sqrt{2}\pi\pm\sqrt{-2\pi^2-\sqrt{2\pi}q-2p\pi}}{2\sqrt{\pi}},
\end{align}
and
\begin{align}\label{especspin2}
\epsilon=\frac{-\sqrt{2}\pi\pm\sqrt{-2\pi^2+\sqrt{2\pi}q-2p\pi}}{2\sqrt{\pi}}.
\end{align}
In equations (\ref{especspin1}) and (\ref{especspin2}), $\pi$ represents one of the solutions of the cubic polynomial according to the Ferrari's method, and explicitly is given by
\begin{equation}\label{pi1}
\pi=\left[-\frac{q_1}{2}+\sqrt{\frac{q_1^2}{4}+\frac{p_1^3}{27}}\right]^{\frac{1}{3}}+\left[-\frac{q_1}{2}-\sqrt{\frac{q_1^2}{4}+\frac{p_1^3}{27}}\right]^{\frac{1}{3}}-\frac{p}{3},
\end{equation}
with $q_1 = \frac{rp}{3}-\frac{q^2}{8}-\frac{p^3}{108}$ and $p_1=-r-\frac{p^2}{12}$. Here it is important to note that the way in which the energy spectra of equations (\ref{especspin1}) and (\ref{especspin2}) are presented does not coincide with that reported in the Ref. \cite{Oliv2}. However, as will be demonstrated later, the results obtained in this section are correct, since they are adequately reduced to those expected for the Dirac oscillator without spin symmetry on flat spacetime.

The radial function $\Phi(r)$ can be obtained from the differential equation \cite{Magh}
\begin{equation}\label{Soledif}
\frac{d^2R(r)_{nml}}{dr^2}+\left(\frac{A_r}{r}+\frac{B_r}{r^2}+C_r\right)R(r)_{nml}=0,
\end{equation}
which has the particular solution
\begin{equation}\label{Hass}
R(r)_{nml}=r^{\frac{1}{2}+\sqrt{\frac{1}{4}-B_r}}e^{-\sqrt{-C_r}r}L_n^{2\sqrt{\frac{1}{4}-B_r}}\left(2\sqrt{-C_r}r\right).
\end{equation}
Thus, by comparing equations (\ref{equdiffac}) and (\ref{Hass}) we obtain
\begin{equation}
A_r=\frac{\beta}{4\delta}+\frac{\epsilon^2-1}{4\delta\alpha^2},\hspace{0.5cm} B_r=-\left[\frac{l\left(l+1\right)+A\left(2\lambda+A+1\right)}{4}-\frac{3}{16}\right], \hspace{0.5cm} C_r=-\frac{1}{4}.
\end{equation}
Therefore, the radial eigenfunction is explicitly given by
\begin{equation}\label{Phisol}
R^1_{nl}=\left(\sqrt{\delta}r\right)^{2\sigma+1}\exp\left(-\frac{\delta{r^2}}{2}\right)L_n^{2\sigma}\left(\delta{r^2}\right),
\end{equation}
where
\begin{equation}\label{sigma}
\sigma=\frac{\sqrt{\left(l+\frac{1}{2}\right)^2+A\left(2\lambda+A+1\right)}}{2}.
\end{equation}
Now, from the equations (\ref{equdiffac}) and (\ref{sigma}) we have the following relation
\begin{equation}
k_s=\frac{1}{2}+\sigma,
\end{equation}
and therefore the radial wave functions for $R^1_{nk_s}$ can be written as
\begin{equation}
R^1_{nk_s}=\left(\sqrt{\delta}r\right)^{2k_s-\frac{1}{2}}\exp\left(-\frac{\delta{r^2}}{2}\right)L_n^{2k_s-1}\left(\delta{r^2}\right).
\end{equation}
The radial wave functions for $R^2_{nk_s}$ can be obtained from equation (\ref{eqdif2}) and explicitly are given by
\begin{align}\nonumber
R^2_{nk_s}=&\frac{\left[-2\delta{rL^{2k_s}_{n-1}}\left(\delta{r^2}\right)+\left(\left(-\delta+\mu\alpha^2\lambda+w\right)r^2+2\gamma+\frac{1}{2}+A+\lambda\right)L^{2k_s-1}_n\left(\delta{r^2}\right)\right]\alpha}{\left(1+\epsilon\right)r}\times\\
&\times\left(\sqrt{\delta}r\right)^{\left(2k_s-\frac{1}{2}\right)}e^{-\frac{\delta{r^2}}{2}}.
\end{align}

\subsection{SU(1,1) radial coherent states in the spin limit}
In this section we will construct the relativistic coherent states for the radial functions $R^1_{nk}$ and $R^2_{nk}$. The $SU(1,1)$ Perelomov coherent states are defined by \cite{Pere}
\begin{equation}\label{defper}
|\varsigma\rangle=D\left(\xi\right)|k,0\rangle=\left(1-|\xi|^2\right)^k\sum^\infty_{n=0}\sqrt{\frac{\Gamma\left(n+2k\right)}{n!\Gamma\left(2k\right)}}\xi^n|k,n\rangle,
\end{equation}
where $D\left(\xi\right)=\exp(\xi K_{+}-\xi^{*}K_{-})$ is the displacement operator. Thus, by substituting the states $R^1_{nk}$ and $R^2_{nk}$ in this expression we obtain
\begin{align}
R^1_{nk_s}\left(r,\xi\right)=&\frac{N}{\delta^{\frac{1}{4}}}\left[\delta\left(1-|\xi|^2\right)\right]^{k_s}r^{2k_s-\frac{1}{2}}e^{\left(-\frac{\delta{r^2}}{2}\right)}\sum^\infty_{n=0}\xi^nL^{2k_s-1}_n\left(\delta{r^2}\right),\\
R^2_{nk_s}\left(r,\xi\right)=&\frac{N\alpha}{\delta^{\frac{1}{4}}\left(1+\epsilon\right)}\left[\delta\left(1-|\xi|^2\right)\right]^{k_s}r^{2k_s-\frac{3}{2}}e^{\left(-\frac{\delta{r^2}}{2}\right)}\left[-2\delta{r}\sum^\infty_{n=0}\xi^nL^{2k_s}_{n-1}\left(\delta{r^2}\right)+\right.\\
&\left.\left[\left(-\delta+\mu\alpha^2\lambda+w\right)r^2+2\gamma+\frac{1}{2}+A+\lambda\right]\sum^\infty_{n=0}\xi^nL^{2k_s-1}_{n}\left(\delta{r^2}\right)\right].
\end{align}
The sums associated with these functions can be calculated using the Laguerre polynomials generating function
\begin{equation}
\sum^\infty_{n=0}L^\nu_n(x)y^n=\frac{e^{{-xy}/\left(1-y\right)}}{\left(1-y\right)^{\nu+1}}.
\end{equation}
Therefore, the radial coherent states $R^1_{nk_s}\left(r,\xi\right)$ and $R^2_{nk_s}\left(r,\xi\right)$ for the modified Dirac oscillator in curved spacetime with the spin limit are given by
\begin{align}
R^1_{nk_s}\left(r,\xi\right)=&\frac{N}{\delta^{\frac{1}{4}}}\left[\frac{\delta\left(1-|\xi|^2\right)}{\left(1-\xi\right)^2}\right]^{k_s}r^{2k-\frac{1}{2}}e^{\frac{\delta{r^2}}{2}\left(\frac{\xi+1}{\xi-1}\right)},\\
R^2_{nk_s}\left(r,\xi\right)=&\frac{N\alpha}{\delta^{\frac{1}{4}}\left(1+\epsilon\right)}\left[\frac{\delta\left(1-|\xi|^2\right)}{\left(1-\xi\right)^2}\right]^{k_s}r^{2k_s-\frac{3}{2}}\left[-2\delta{r}e^{\frac{\delta{r^2}}{2}\left(\frac{\xi+1}{\xi-1}\right)}+\right.\\
&\left.\left[\left(-\delta+\mu\alpha^2\lambda+w\right)r^2+2\gamma+\frac{1}{2}+A+\lambda\right]e^{\frac{\delta{r^2}}{2}\left(\frac{\xi+1}{\xi-1}\right)}\right].
\end{align}

\section{The pseudospin limit $\Sigma(r)=0, V(r)=-U(r), \Sigma(r)=-2U(r)$}
Now we will consider the pseudospin limit case. In this scenario, the sum between the scalar and vector potentials is zero $\Sigma(r)=V(r)+S(r)=0$, that is $V(r)=-U(r), \Sigma(r)=-2U(r)$.
Under this limit, from equations (\ref{eqdif1}) and (\ref{eqdif2}), we can write the decoupled differential equation for $R_{2\lambda}(r)$ as
\begin{equation}\label{difpseu2}
\left[\frac{d^2}{dr^2}-\frac{\lambda\left(\lambda-1\right)+A\left(2\lambda+A-1\right)}{r^2}-\delta^2r^2-\beta+\frac{\epsilon^2-1}{\alpha^2}\right]R_{2\lambda}(r)=0.
\end{equation}
Following a procedure similar to that used in the case of the spin limit, from the substitutions $x={\delta}r^2$ and $R_{2\lambda}(r)=x^{\frac{1}{4}}\Psi$, the differential equation (\ref{difpseu2}) can be written as
\begin{equation}\label{equdifpseu}
\left[-x\frac{d^2}{dx^2}+\left(\frac{\bar{l}\left(\bar{l}+1\right)+A\left(2\lambda+A-1\right)}{4}-\frac{3}{16}\right)\frac{1}{x}+\frac{x}{4}\right]\Psi=\left(\frac{\epsilon^2-1}{4\delta\alpha^2}-\frac{\beta}{4\delta}\right)\Psi,
\end{equation}
where
\begin{align}\label{difpseudo}\nonumber
\lambda\left(\lambda-1\right)=&\bar{l}\left(\bar{l}+1\right),\hspace{0.5cm}\beta=w+2{\lambda}w+2wA-\lambda\alpha^2\mu-2\lambda^2\alpha^2\mu-2\lambda\alpha^2{\mu}A,\\
\delta=&\sqrt{\left(w-\alpha^2\lambda\mu\right)^2+2\mu+2\mu\epsilon}.
\end{align}
In order to apply the Schr\"odinger factorization, we expand equation (\ref{factosch}) and comparing it with the differential equation for $R_{2\lambda}(r)$ (\ref{equdifpseu}), the following relations are obtained
\begin{align}
a=&\pm\frac{1}{2},\hspace{0.5cm} c=\pm\frac{1}{2}, \hspace{0.5cm} f=\mp\left(\frac{\epsilon^2-1}{4\delta\alpha^2}-\frac{\beta}{4\delta}\right),\hspace{0.5cm}b=\mp\left(\frac{\epsilon^2-1}{4\delta\alpha^2}-\frac{\beta}{4\delta}\right)-1,\\
g=&\left[-\frac{\bar{l}\left(\bar{l}+1\right)+A\left(2\lambda+A-1\right)}{4}-\frac{3}{16}\right]+\left[\mp\left(\frac{\epsilon^2-1}{4\delta\alpha^2}-\frac{\beta}{4\delta}\right)-1\right]\left[\mp\left(\frac{\epsilon^2-1}{4\delta\alpha^2}-\frac{\beta}{4\delta}\right)\right].
\end{align}
Thus, the factorization of the differential equation (\ref{equdifpseu}) is
\begin{equation}
\left[\mathscr{X}_{\mp}\mp1\right]\mathscr{X}_{\pm}=\bar{\Pi}+\bar{\Gamma}\left(\bar{\Gamma}\pm1\right),
\end{equation}
where
\begin{equation}
\bar{\Pi}=\left[-\frac{\bar{l}\left(\bar{l}+1\right)+A\left(2\lambda+A-1\right)}{4}-\frac{3}{16}\right], \hspace{0.5cm}\bar{\Gamma}=\left(\frac{\epsilon^2-1}{4\delta\alpha^2}-\frac{\beta}{4\delta}\right).
\end{equation}
The Schr\"odinger operators in the pseudospin limit are
\begin{align}\label{operschps}
\mathscr{X}_{\pm}=\mp x\frac{d}{dx}+\frac{x}{2}-\left(\frac{\beta}{4\delta}-\frac{\epsilon^2-1}{4\delta\alpha^2}\right),
\end{align}
From equations (\ref{equdifpseu}) and (\ref{operschps}) we can write the operators
\begin{align}\label{operschps2}
\mathscr{P}_{\pm}=\mp x\frac{d}{dx}+\frac{x}{2}-\mathscr{E}_3,
\end{align}
where the operator $\mathscr{E}_3$ is explicitly given by
\begin{equation}\label{oper3pseud}
\mathscr{E}_3=\left[-x\frac{d^2}{dx^2}+\left(\frac{\bar{l}\left(\bar{l}+1\right)+A\left(2\lambda+A-1\right)}{4}-\frac{3}{16}\right)\frac{1}{x}+\frac{x}{4}\right].
\end{equation}
Here, just like in the case of the spin limit, the operators $\mathscr{P}_{\pm}$ and $\mathscr{E}_3$ close the $su(1,1)$ algebra, i.e., they satisfy the following commutators
\begin{equation}
\left[\mathscr{E}_3, \mathscr{P}_{\pm}\right]=\mp\mathscr{P}_{\pm},\hspace{0.5cm} \left[\mathscr{P}_-,\mathscr{P}_+\right]=2\mathscr{E}_3.
\end{equation}

The energy spectrum is obtained in a similar way to the spin limit, i.e. by means of equations (\ref{tercerop}) and (\ref{casi}). Hence, from equation (\ref{casi}) we obtain that group number $\bar{k}_{Ps}$ is given by
\begin{align}\label{opk}
\bar{k}_{Ps}=\frac{1}{2}\pm\frac{\sqrt{1+4A^2+8A\lambda+4A+4\bar{l}+4\bar{l}^2}}{4}.
\end{align}
Therefore, from equations (\ref{OPER3}) and (\ref{casi}) we can write
\begin{equation}\label{ec1ener}
\bar{k}_{Ps}+\bar{n}=\left[\frac{\epsilon^2-1}{4\delta\alpha^2}-\frac{\beta}{4\delta}\right].
\end{equation}
As it was shown in Section $2$, the energy spectrum can be computed from this expression finding the roots of the following
polynomial
\begin{equation}
\epsilon^4+p\epsilon^2+q\epsilon+r=0,
\end{equation}
where in the pseudospin limit we have that
\begin{align}\label{var1ps}
p=&2a, \hspace{1.0cm} q=\left(-2{\mu}b^2\right),\hspace{1.0cm} r=a^2-b^2c,\\\label{var2ps}
a=&-1-\beta\alpha^2, \hspace{1.0cm} b=4\alpha^2\left(k_{Ps}+n\right),\hspace{1.0cm} c=\left(-2\mu+\left(w-\alpha^2\lambda\mu\right)^2\right),
\end{align}
The energy spectrum can be calculated using equations (\ref{especspin1}), (\ref{especspin2}) and (\ref{pi1}). It's important to note that in this limit, the values for $a$, $b$, $c$ and $b$ are different and are given by equations (\ref{var1ps}) and (\ref{var2ps}).

Following a similar procedure as in Section $2$, we will obtain the radial wave functions for the pseudospin limit with the help of equations (\ref{Soledif}), (\ref{Hass}) and (\ref{equdifpseu})
\begin{equation}\label{Phisol2}
\Psi_{\bar{n}\bar{l}}=\left(\sqrt{\delta}r\right)^{2\bar{\varrho}+1}\exp\left(-\frac{\delta{r^2}}{2}\right)L_n^{2\varrho}\left(\delta{r^2}\right).
\end{equation}
where
\begin{equation}\label{varrho}
\bar{\varrho}=\frac{\sqrt{\left(\bar{l}+\frac{1}{2}\right)^2+A\left(2\lambda+A+1\right)}}{2},
\end{equation}
Thus, from equation (\ref{opk}) and (\ref{varrho}) we have the following relation
\begin{equation}
\bar{k}_{Ps}=\frac{1}{2}+\bar{\varrho}.
\end{equation}

Therefore the explicit form of the radial wave functions $R^1_{nk_s}$ and $R^2_{nk_s}$ is obtained from equations (\ref{eqdif1}), (\ref{eqdif2}), (\ref{Soledif}) and (\ref{Hass})
\begin{align}\label{r1p}
R^1_{\bar{n}\bar{k}}=&\left(\sqrt{\delta}r\right)^{2\bar{k}_{Ps}-\frac{1}{2}}\exp\left(-\frac{\delta{r^2}}{2}\right)L_n^{2\bar{k}_{Ps}-1}\left(\delta{r^2}\right),\\\label{r2p}
R^2_{\bar{n}\bar{k}_{Ps}}=&\frac{\left[-2\delta{rL^{2\bar{k}_{Ps}}_{n-1}}\left(\delta{r^2}\right)+\left(\left(-\delta+\mu\alpha^2\lambda+w\right)r^2+2\gamma+\frac{1}{2}+A+\lambda\right)L^{2\bar{k}_{Ps}-1}_n\left(\delta{r^2}\right)\right]\alpha}{\left(1+\epsilon\right)r}\times\nonumber\\
&\times\left(\sqrt{\delta}r\right)^{\left(2\bar{k}_{Ps}-\frac{1}{2}\right)}e^{-\frac{\delta{r^2}}{2}}.
\end{align}

\subsection{SU(1,1) radial coherent states in the pseudospin limit}
The coherent states for the radial eigenfunctions in the pseudospin limit can be computed following a mathematical procedure similar to that employed in Section $3$. Thus, by substituting the expressions (\ref{r1p}) and (\ref{r2p}) into the equation (\ref{defper}) we obtain that the coherent states $R^1_{\bar{n}\bar{k}_{Ps}}\left(r,\xi\right)$ and $R^2_{\bar{n}{k}_{Ps}}\left(r,\xi\right)$ are given by
\begin{align}
R^1_{\bar{n}\bar{k}_{Ps}}\left(r,\xi\right)=&\frac{N}{\delta^{\frac{1}{4}}}\left[\frac{\delta\left(1-|\xi|^2\right)}{\left(1-\xi\right)^2}\right]^{\bar{k}_{Ps}}r^{2\bar{k}_{Ps}-\frac{1}{2}}e^{\frac{\delta{r^2}}{2}\left(\frac{\xi+1}{\xi-1}\right)},\\\nonumber
R^2_{\bar{n}\bar{k}_{Ps}}\left(r,\xi\right)=&\frac{N\alpha}{\delta^{\frac{1}{4}}\left(1+\epsilon\right)}\left[\frac{\delta\left(1-|\xi|^2\right)}{\left(1-\xi\right)^2}\right]^{\bar{k}_{Ps}}r^{2\bar{k}_{Ps}-\frac{3}{2}}\left[-2\delta{r}e^{\frac{\delta{r^2}}{2}\left(\frac{\xi+1}{\xi-1}\right)}+\right.\\
&\left.\left[\left(-\delta+\mu\alpha^2\lambda+w\right)r^2+2\gamma+\frac{1}{2}+A+\lambda\right]e^{\frac{\delta{r^2}}{2}\left(\frac{\xi+1}{\xi-1}\right)}\right].
\end{align}
Note that to obtain these expressions the generating function of the Laguerre polynomials has been used.

\section{Time evolution of the coherent states in the spin and pseudospin limit}

In this Section we will compute the time evolution of the coherent states for the modified Dirac oscillator in curved spacetime with in the spin and pseudospin limit. To this end, we can
write the equation (\ref{equdiffac}) as
\begin{equation}\label{Hgam}
H_x\Phi=\left(\frac{\beta}{4\delta}+\frac{\epsilon^2-1}{4\delta\alpha^2}\right)\Phi,
\end{equation}
where
\begin{equation}\label{second4}
H_x=\left[-x\frac{d^2}{dx^2}+\left(\frac{l\left(l+1\right)+A\left(2\lambda+A+1\right)}{4}-\frac{3}{16}\right)\frac{1}{x}+\frac{x}{4}\right].
\end{equation}
From equations (\ref{OPER3}) and (\ref{Hgam}) it follows that
\begin{equation}\label{OPB3ET}
H_x\Phi=\mathscr{B}_3\Phi=\left(\frac{\beta}{4\delta}+\frac{\epsilon^2-1}{4\delta\alpha^2}\right)\Phi.
\end{equation}
Thus, from the action of the operator $\mathscr{B}_3$ on the $SU(1,1)$ states (see equation (\ref{tercerop})) we again arrive to equation (\ref{ecuener})
\begin{equation}
\epsilon^4+p\epsilon^2+q\epsilon+r=0,
\end{equation}
from which we can obtain the energy spectrum following the procedure explained in subsection $2.1$.

The time evolution operator for an arbitrary Hamiltonian is defined as \cite{Cohen}
\begin{equation}\label{UTEM}
\mathfrak{U}(\tau)=e^{-iH_{x}\tau/\hbar}=e^{-i\mathscr{B}_3\tau/\hbar},
\end{equation}
where $\tau$ represents a fictitious time. The time evolution of the Perelomov coherent states can be written as \cite{Mann,Gerry,NOS1}
\begin{equation}\label{PERET}
|\zeta(\tau)\rangle =\mathfrak{U}(\tau)|\zeta\rangle=\mathfrak{U}(\tau)D(\xi)\mathfrak{U}^\dag(\tau)\mathfrak{U}(\tau)|k,0\rangle,
\end{equation}
where $D(\xi)=\exp(\xi \mathscr{T}_{+}-\xi^{*}\mathscr{T}_{-})$ is the displacement operator and $\xi$ is a complex number.
Thus, the time evolution of the lowest normalized $|k,0\rangle$ using equation (\ref{tercerop}) is therefore
\begin{equation}\label{evest1}
\mathfrak{U}(\tau)|k,0\rangle=e^{-ik\tau/\hbar}|k,0\rangle.
\end{equation}

We will now calculate the time evolution of the raising and lowering operators by using the BCH (Baker-Campbell-Hausdorff) formula as follows
\begin{align}
\mathscr{T}_+(\tau)=&\mathfrak{U}^\dag(\tau)\mathscr{T}_+\mathfrak{U}(\tau)=\mathscr{T}_+e^{i\tau/\hbar},\\
\mathscr{T}_-(\tau)=&\mathfrak{U}^\dag(\tau)\mathscr{T}_-\mathfrak{U}(\tau)=\mathscr{T}_-e^{-i\tau/\hbar}.
\end{align}
Therefore, the similarity equation involved in equation (\ref{PERET}), is given by
\begin{equation}\label{opd1}
\mathfrak{U}(\tau)D(\xi)\mathfrak{U}^\dag(\tau)=e^{\xi \mathscr{T}_+(-\tau)-\xi^*\mathscr{T}_-(-\tau)}=e^{\xi(-\tau)\mathscr{T}_+ - \xi(-\tau)^*\mathscr{T}_-},
\end{equation}
were $\xi(t)=\xi e^{\frac{-i\tau}{\hbar}}$. Similarly, we can obtain an expression for the time evolution of the displacement operator in its normal form by defining $\zeta(t)=\zeta e^{\frac{-i\tau}{\hbar}}$ as
\begin{equation}\label{opedes}
D(\xi(t))=e^{\zeta(t) \mathscr{T}_+}e^{\eta \mathscr{T}_0}e^{-\zeta(t)^*\mathscr{T}_-}.
\end{equation}
Hence, from equations (\ref{evest1}) and (\ref{opedes}) we obtain that the time dependent Perelomov coherent state is
\begin{equation}
|\zeta(t)\rangle=e^{- k\tau/\hbar}e^{\zeta(-\tau)\mathscr{T}_+}e^{\eta \mathscr{T}_0}e^{-\zeta(-\tau)^*\mathscr{T}_-}|k,0\rangle.
\end{equation}

Therefore, from this expression we can compute the time evolution of the radial coherent states for the modified Dirac oscillator in curved spacetime with the spin limit
\begin{align}
R^1_{nk_S}\left(r,\xi\right)=&\frac{N}{\delta^{\frac{1}{4}}}\left[\frac{\delta\left(1-|\xi|^2\right)}{\left(1-\xi\right)^2}\right]^{k_s}r^{2k-\frac{1}{2}}e^{-ik_{s}}e^{\frac{\delta{r^2}}{2}\left(\frac{{\xi}e^{i\tau/\hbar}+1}{{\xi}e^{i\tau/\hbar}-1}\right)},\\\nonumber
R^2_{nk_S}\left(r,\xi\right)=&\frac{N\alpha}{\delta^{\frac{1}{4}}\left(1+\epsilon\right)}\left[\frac{\delta\left(1-|\xi|^2\right)}{\left(1-\xi\right)^2}\right]^{k_s}r^{2k_s-\frac{3}{2}}e^{-ik_{s}}e^{\frac{\delta{r^2}}{2}\left(\frac{{\xi}e^{i\tau/\hbar}+1}{{\xi}e^{i\tau/\hbar}-1}\right)}\times\\
&\left[-2\delta{r}+\left(-\delta+\mu\alpha^2\lambda+w\right)r^2+2\gamma+\frac{1}{2}+A+\lambda\right].
\end{align}
Analogously, it can be shown that the time evolution of the radial coherent states with the pseudospin limit are given by
\begin{align}
R^1_{\bar{n}\bar{k}_{Ps}}\left(r,\xi\right)=&\frac{N}{\delta^{\frac{1}{4}}}\left[\frac{\delta\left(1-|\xi|^2\right)}{\left(1-\xi\right)^2}\right]^{\bar{k}_{Ps}}r^{2k-\frac{1}{2}}e^{-ik_{s}}e^{\frac{\delta{r^2}}{2}\left(\frac{{\xi}e^{i\tau/\hbar}+1}{{\xi}e^{i\tau/\hbar}-1}\right)},\\\nonumber
R^2_{\bar{n}\bar{k}_{Ps}}\left(r,\xi\right)=&\frac{N\alpha}{\delta^{\frac{1}{4}}\left(1+\epsilon\right)}\left[\frac{\delta\left(1-|\xi|^2\right)}{\left(1-\xi\right)^2}\right]^{\bar{k}_{Ps}}r^{2\bar{k}_{Ps}-\frac{3}{2}}e^{-ik_{s}}e^{\frac{\delta{r^2}}{2}\left(\frac{{\xi}e^{i\tau/\hbar}+1}{{\xi}e^{i\tau/\hbar}-1}\right)}\times\\
&\left[-2\delta{r}+\left(-\delta+\mu\alpha^2\lambda+w\right)r^2+2\gamma+\frac{1}{2}+A+\lambda\right].
\end{align}

\subsection{Energy spectrum for the Dirac oscillator without spin symmetry on flat spacetime}
Now, we will analyze the energy spectrum in the case of the Dirac oscillator without spin symmetry on flat spacetime. If we make $A=0$ and $\mu=0$, the equation (\ref{ec1ener}) can be written as
\begin{equation}\label{esfmuo}
\frac{1}{2}+\frac{1+2l}{4}+n=\left[\frac{1-2{\lambda}}{4}+\frac{\epsilon^2-1}{4w\alpha^2}\right].
\end{equation}
Therefore, the energy spectrum for Dirac oscillator without spin symmetry on flat spacetime is explicitly given by
\begin{equation}\label{epplan}
\epsilon=\sqrt{1+2w\alpha^2\left(2n+\lambda+l+1\right)}.
\end{equation}
This agrees with the results reported in \cite{Oliv2} and \cite{Alha3} for the case of the Dirac oscillator without spin symmetry on flat space-time. However, the way to obtain the energy spectrum given in equation (\ref{epplan}) directly from equations (\ref{especspin1}) and (\ref{especspin2}) is not straightforward. Another way to return to the energy spectrum in the case of the Dirac oscillator without spin symmetry on flat spacetime is to make $\mu=0$($q=0$) and $A=0$ in equation (\ref{ecuener}), that is,
\begin{equation}\label{ecuenerSM}
\epsilon^4+p\epsilon^2+r=0.
\end{equation}
This expression is easy to solve without using Ferrari's solution, obtaining the following four roots
\begin{align}\nonumber
\epsilon_1&=\sqrt{-a+\sqrt{a^2-r}}, \hspace{0.5cm}\epsilon_2=-\sqrt{-a+\sqrt{a^2-r}},\\\epsilon_3&=\sqrt{-a-\sqrt{a^2-r}},\hspace{0.5cm} \epsilon_4=-\sqrt{-a-\sqrt{a^2-r}}.
\end{align}
If we substitute the value of $a$ and $r$ given in equation (\ref{var1}), we obtain that only the roots $\epsilon_1$ and $\epsilon_2$ return us to the energy spectrum of the Dirac oscillator without spin symmetry on flat spacetime. This fact was also observed and commented on the Ref. \cite{Oliv2}.

\section{Concluding Remarks}

In this work, we studied the problem of the modified Dirac oscillator in curved spacetime with spin and pseudospin symmetries from an algebraic approach. We applied the Sch\"odinger factorization to the uncoupled second-order radial equations in the spin and pseudospin limits to obtain the $su(1,1)$ algebraic generators. The energy spectrum for the spin and pseudospin limits were obtained by using the theory of unitary representations of the $su(1,1)$ Lie algebra. We found with our method the energy spectrum in the limits $\mu=0$ and $A = 0$, which is consistent with the spectrum of the Dirac oscillator without spin symmetry on flat space-time, as expected. The eigenfunctions for each limit were obtained in a closed form in terms of the Laguerre polynomials.

Finally, we computed the $SU(1,1)$ radial coherent states and their time evolution in the spin and pseudospin limit for each spinor component.

\end{document}